\documentclass{optica-article}

\journal{optica}

\articletype{Research Article}

\usepackage{lineno}

\begin{document}

\title{Micropascal-sensitivity ultrasound sensors based on optical microcavities}

\author{Hao Yang,\authormark{1,2,5} Xuening Cao,\authormark{1,2,5} Zhi-Gang Hu,\authormark{1,2} Yimeng Gao,\authormark{1,2} Yuechen Lei,\authormark{1,2} Min Wang,\authormark{1,2} Zhanchun Zuo,\authormark{1,3} Xiulai Xu\authormark{4} and Bei-Bei Li\authormark{1,3,*}}

\address{\authormark{1}Beijing National Laboratory for Condensed Matter Physics, Institute of Physics, Chinese Academy of Sciences, Beijing 100190, China\\
\authormark{2}University of Chinese Academy of Sciences, Beijing 100049, China\\
\authormark{3}Songshan Lake Materials Laboratory, Dongguan 523808, Guangdong, China\\
\authormark{4}State Key Laboratory for Mesoscopic Physics and Frontiers Science Center for Nano-optoelectronics, School of Physics, Peking University, 100871 Beijing, China\\
\authormark{5}The authors contributed equally to this work.}
\email{\authormark{*}libeibei@iphy.ac.cn}



\begin{abstract}
Whispering gallery mode (WGM) microcavities have been widely used for high-sensitivity ultrasound detection, due to their optical and mechanical resonances enhanced sensitivity. The ultrasound sensitivity of the cavity optomechanical system is fundamentally limited by the thermal noise. In this work, we theoretically and experimentally investigate the thermal-noise-limited sensitivity of a WGM microdisk ultrasound sensor, and optimize the sensitivity by varying the radius and thickness of the microdisk, as well as using a trench structure around the disk. Using a microdisk with a radius of 300~\textmu m and thickness of 2~\textmu m, a peak sensitivity of 1.18~\textmu Pa Hz$^{-1/2}$ is achieved at 82.6~kHz, which is to our knowledge the record sensitivity among the cavity optomechanical ultrasound sensors. Such high sensitivity can improve the detection range of air-coupled ultrasound sensing technology.
\end{abstract}

\section{Introduction}

Ultrasound has played an important role in various applications. For instance, non-destructive testing with ultrasound has been widely used in industrial quality inspection \cite{industrial}. Imaging using ultrasound can detect deeper tissue structures than light and help in medical diagnosis \cite{medical}. Ultrasound can also be applied to positioning and ranging systems\cite{positioning}, which can work in bad weather. Meanwhile, ultrasound detection technology is also facing more challenges. Ultrasound at the interface of different media generally has a large loss due to the acoustic impedance mismatch. Therefore detection usually requires an acoustic impedance-matching coupling agent to achieve high sensitivity. However, for some detection scenarios where coupling agents cannot be used (such as wounds, high temperatures, materials that cannot be damaged, etc.), the air-coupled ultrasound detection is the best choice. Air-coupled ultrasound sensors with higher sensitivity are required to reduce the effect of airborne losses. Commercially used today are usually piezoelectric ultrasound transducers, which are manufactured in a mature process and are easy to integrate with circuits. However, the size of the piezoelectric transducer is generally in the millimeter to centimeter range to meet the requirement of high sensitivity, which greatly reduces the spatial resolution for imaging. 

To overcome these shortcomings, optical ultrasound sensors are developed, with both high sensitivities and spatial resolution, such as Fabry-Perot (F-P) interferometers \cite{FP1,FP2,plano-concave,buckled-dome}, WGM microcavities \cite{microsphere1,microsphere2,microsphere3,microsphere4,microring1,microring2,microring3,microring4,microring5,microbubble1,microbubble2,microbubble3,microbubble4,microdisk,microtoroid}, and Bragg gratings \cite{silicon-on-insulator,Bragg}. The pressure changes created by the incoming ultrasound cause optical resonance shift in these resonators and can be optically read out with high sensitivity. F-P cavity ultrasound sensors made of graphene \cite{FP1} and silver films \cite{FP2} on optical fibers can already achieve tens of micropascals sensitivity. However, millimeter-scale dimensions are required to achieve such high sensitivities. Recently, a new F-P cavity has been designed, which comprises a solid plano-concave polymer microcavity \cite{plano-concave} formed between two highly reflective mirrors. It has realized an equivalent noise pressure of 1.6~mPa Hz$^{-1/2}$, wide directivity, and its application in biomedical imaging has been demonstrated. Owing to their ultrahigh quality ($Q$) factors enhanced light-matter interactions, WGM microcavities have demonstrated exceptional performance in ultrasound detection. Various structures have been used for ultrasound sensing, such as microspheres \cite{microsphere1,microsphere2,microsphere3,microsphere4}, microrings \cite{microring1,microring2,microring3,microring4,microring5}, microbubbles \cite{microbubble1,microbubble2,microbubble3,microbubble4}, microdisks \cite{microdisk} and microtoroids \cite{microtoroid}. Among them, cavity optomechanical systems \cite{optomechanics,CavityOptomechanic,APR2014,magnetometry2018} provide an ideal platform for ultrasound sensing due to their dual-resonance enhanced sensitivity. The sensitivity of optomechanical ultrasound sensors is only fundamentally limited by the thermal noise. Therefore reaching the thermal-noise-limited regime is crucial to achieve a better sensitivity. Recently, thermal-noise-limited ultrasound sensitivities at the order of micropascals have been realized \cite{microdisk,microtoroid,buckled-dome}. How to further improve the thermal-noise-limited ultrasound sensitivity still requires a more systematic study.

Our previous work \cite{microtoroid} has theoretically studied the thermal-noise-limited ultrasound sensitivity, without considering the effects of the pressure difference between the upper and lower surfaces of the microdisk, and the spatial overlap between the mechanical mode and ultrasound. In this work, we perform a more systematic study on the thermal-noise-limited ultrasound sensitivity using suspended WGM microdisks both theoretically and experimentally. Our study shows that a trench structure around the microdisk can increase the pressure difference to enhance its response to ultrasound. We study the trends of sensitivities with the radius and thickness of the microdisk, taking into account the pressure difference and spatial overlap. We experimentally fabricate microdisks of different radii and thicknesses with trench structure, and measure their sensitivities in the air in frequency ranges from tens of kHz to more than 1~MHz. We achieve to our knowledge by far the best cavity optomechanical ultrasound sensitivity of 1.18~\textmu Pa Hz$^{-1/2}$ at 82.6~kHz, which corresponds to the second-order flapping mode of a microdisk with a radius of 300~\textmu m and a thickness of 2~\textmu m. Such a high sensitivity can increase the detection range of air-coupled ultrasound sensing technology, which is particularly useful for applications such as positioning systems and gas photoacoustic spectroscopy \cite{trace-gas}.

\section{Theoretical analysis}

\begin{figure}[h!]
\centering\includegraphics[width=10cm]{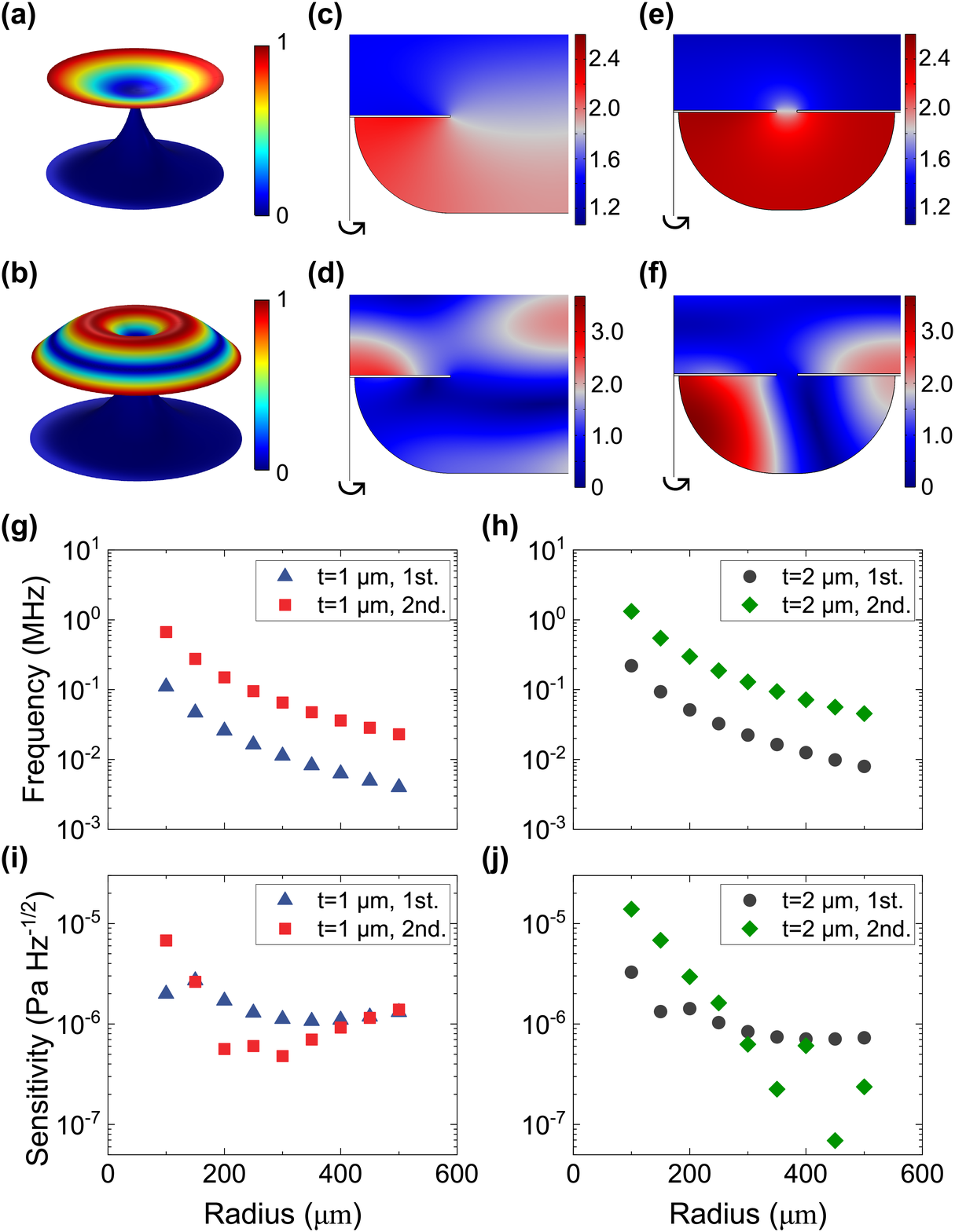}
\caption{(a) and (b) Simulated displacement distributions of the first- and second-order flapping modes, respectively, normalized to their maximum values. (c)-(f) Simulated pressure distributions of the microdisk without ((c) and (d)) and with ((e) and (f)) the trench structure, at the frequencies of the first-order ((c) and (e)) and second-order ((d) and (f)) flapping modes. In (c)-(f), the ultrasonic wave is perpendicularly incident to the microdisk from above. The microdisk used here has a radius of 100~\textmu m and a thickness of 2~\textmu m. The pressure differences are 0.68 (c), 1.91 (d), 1.35 (e), 1.97 (f), respectively. (g) and (h) Simulated mechanical resonance frequencies of the flapping modes of 1-\textmu m-thick and 2-\textmu m-thick microdisks respectively, as a function of the disk radius. (i) and (j) Calculated sensitivities of microdisks, with thicknesses of 1~\textmu m and 2~\textmu m, respectively. The blue triangles and red squares represent the results of the first- and second-order flapping modes of the microdisk with a thickness of 1~\textmu m. The black circles and green rhombuses represent the results of the first- and second-order flapping modes of the microdisk with a thickness of 2~\textmu m.}
\end{figure}

The ultrasound sensitivity is limited by the noise of the sensor. In an optomechanical system, the main noise sources include the thermal-mechanical noise, optical shot noise and backaction noise from the probe light \cite{magnetometry2018,Nanophotonics2021,optomechanics}. As studied in previous works, for a proper choice of the parameters in the detection system, including optical power, optical $Q$ factor, mechanical $Q$ factor, and optomechanical coupling coefficient $G=d\omega/dx$ (with $\omega$ and $x$ denoting the optical resonance angular frequency and mechanical displacement), the measurement strength can be made strong enough to enter a thermal-noise-dominant regime. This is beneficial to achieving a better sensitivity. In this work, we focus on the study of the thermal-noise-limited ultrasound sensitivity and optimize the sensitivity by varying the geometric parameters of a microdisk. The thermal-noise-limited ultrasound sensitivity $P_{\rm{min}}$ can be expressed as Eq. (1) \cite{optomechanics,microdisk}

\begin{align}
P_{\rm{min}}(\omega)= \frac{1}{r\zeta A}\sqrt{2m\gamma k_{\rm{B}} T},
\end{align}
where \emph{r} is the ratio of the pressure difference between the upper and lower surfaces of the sensor to the peak pressure at the antinode of the incident ultrasonic wave, and $\zeta$ is the spatial overlap between the incident ultrasound and the mechanical displacement profile of the sensor. \emph{A} is the sensor area. The square-root term quantifies the thermal-mechanical force spectral density, introduced by both the intrinsic damping of the mechanical resonator and collisions with the gas molecules around the sensor. Here \emph{m} and $\gamma$ are the effective mass and damping rate of the mechanical mode of the sensor, $k_\mathrm{B}$ is the Boltzmann constant, and $T$ is the temperature.

From this equation we can see how the parameters affect the sensitivity. First, a higher mechanical $Q$ factor (smaller $\gamma$) is helpful for achieving a better sensitivity. Using a microdisk with a thin silicon pedestal can decrease the clamping loss, and therefore increase the mechanical $Q$ factors. Second, a better sensitivity can be achieved by using mechanical modes with larger spatial overlaps $\zeta$. For a microdisk ultrasound sensor, the flapping modes have good spatial overlaps with the ultrasound. Therefore in this work we mainly consider the flapping modes of the microdisk. Figures 1(a) and 1(b) show the normalized displacement distributions of the first- and second-order flapping modes of a microdisk (radius of 100~\textmu m, thickness of 2~\textmu m, and pedestal radius of 10~\textmu m), obtained from the finite element method (FEM) simulations. Their spatial overlaps with the ultrasound perpendicularly incident to the microdisk are around 0.58 and 0.23, respectively. The spatial overlap is smaller for the second-order flapping mode. This is because the spatial overlap is related to the direction of the mechanical displacement, and the displacement cancellation in opposite directions of the second-order flapping mode decreases its spatial overlap. Their mechanical resonance frequencies are 219~kHz and 1.32~MHz, respectively.

Pressure difference is also an important factor affecting sensitivity, and a larger pressure difference $r$ gives a better sensitivity. The pressure difference $r$ depends on the ultrasound frequency and the structure of the sensor. In our work we find that the pressure difference at low frequencies can be increased by using a trench structure around the microdisk sensor. We use a two-dimensional axial symmetry model to simulate the pressure distribution of the microdisk, with the axis of symmetry located at the center of the microdisk. Figures 1(c) and 1(d) represent the pressure distributions of the microdisk without a trench structure, at the first- and second-order flapping modes, with the pressure difference $r=$0.68 and $r=$1.91, respectively. The pressure difference is higher for the second-order flapping mode, due to the increasing spatial gradient of the pressure wave at higher frequencies. Figures 1(e) and 1(f) show the pressure distributions of the microdisk with a trench structure, at the first- and second-order flapping modes, with the pressure difference $r=$1.35 and $r=$1.97, respectively. It can be seen that the pressure difference of the first-order flapping mode is significantly increased by the trench structure, while that of the second-order flapping mode is not affected much. This is because, at lower frequencies, the trench structure can greatly increase the ultrasonic wave reflections between the substrate and the microdisk, and therefore increase the pressure difference. While for the second-order flapping mode, the reflection between the substrate and the microdisk is decreased, due to the more pronounced diffraction at the trench structure for higher-frequency (shorter-wavelength) acoustic waves.

Equation (1) also suggests that the sensitivity gets better for a larger sensor area $A$, but the effect of the pressure difference also needs to be considered. In the following we theoretically study the ultrasound sensitivity at the first- and second-order flapping modes of microdisk sensors with different radii, taking into account the pressure difference $r$ and spatial overlap $\zeta$. Figures 1(g) and 1(h) show the simulated resonance frequencies of the first- and second-order flapping modes of the microdisk, as a function of the disk radius, with thicknesses of 1~\textmu m and 2~\textmu m, respectively. It can be seen that the resonance frequency decreases with the increase of radius and the decrease of thickness. We obtain the spatial overlap and pressure difference through simulation, and obtain the corresponding sensitivities for microdisks with different radii and thicknesses of 1~\textmu m and 2~\textmu m, as shown in Figs. 1(i) and 1(j), respectively. According to Eq. (1), on the one hand, the increase of the radius will increase the sensor area and thus improve the sensitivity. On the other hand, the decrease of the resonance frequency will hinder the improvement of the sensitivity due to the decreased pressure difference. Therefore, under the effect of these two factors, the sensitivity gets improved first and then degrades. By comparing the first- and second-order flapping modes, it can be seen that the sensitivities of the first-order flapping mode are better for the small-radius microdisks because it has a lager spatial overlap. However, when the radius becomes lager, the effect of the pressure difference increases, resulting in worse sensitivities of the first-order flapping mode.

\section{Device fabrication and measurement}

\begin{figure}[h!]
\centering\includegraphics[width=10cm]{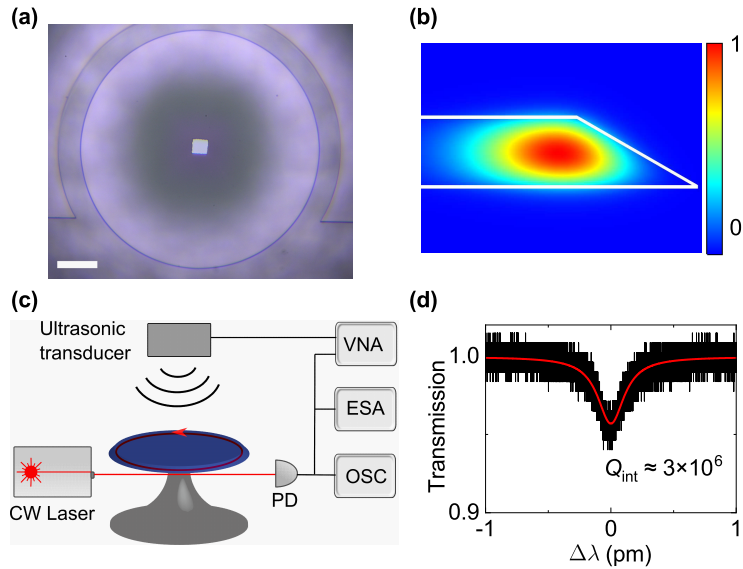}
\caption{(a) Top-view optical microscope image of a microdisk with a trench structure. The scale bar corresponds to 50~\textmu m. (b) FEM simulated optical field distribution of the fundamental WGM mode of the microdisk. (c) Experimental setup to measure the microdisk response to ultrasound. PD, photodetector; VNA, vector network analyzer; OSC, oscillosope; ESA, electronic spectrum analyzer. (d) Optical transmission spectrum of the microdisk, with the intrinsic $Q$ factor of 3 million.}
\end{figure}

For experimental demonstration, we fabricate a series of microdisks of various radii and thicknesses, with trench structure around the microdisks. The fabrication of the microdisks includes photolithography, hydrofluoric acid wet etching, and xenon difluoride (XeF$_2$) dry etching processes. During the XeF$_2$ dry etching process, the silicon pedestal under the silica microdisk becomes thinner to improve mechanical compliance. Meanwhile, a trench is etched around the microdisk while the rest of the silicon substrate is preserved due to our special design. Figure 2(a) shows an optical microscopic image of the microdisk with a radius of 150~\textmu m. We use a 1~\textmu m diameter tapered fiber in the notch of the trench to couple light into the WGM of the microdisk. The cross-section electric field distribution of the fundamental mode close to the surface of the microdisk is shown in Fig. 2(b). The optical field is localized at the periphery of the microdisk and can couple to the tapered fiber evanescently.

Characterization of the microdisk ultrasound sensors is carried out using the setup shown in Fig. 2(c). A continuous wave tunable laser in the 1550~nm band is used to couple light into the microdisk, and the transmitted light from the tapered fiber is collected by a photodector, and then measured by an oscilloscope, electronic spectrum analyzer (ESA), and vector network analyzer (VNA), respectively. The mechanical and optical modes of the microdisk as well as its ultrasound response can be obtained by analyzing the transmitted light. An oscilloscope is used to read out the optical transmission spectrum of the microdisk when the wavelength of the laser is scanned. From the optical transmission spectrum, we can obtain the optical $Q$ factors of microdisks, which are at the 10$^6$ level. Figure 2(d) shows a typical under-coupled transmission spectrum with an intrinsic optical $Q$ factor of 3 million. In the experiments of ultrasound detection, it is necessary to lock the laser wavelength at the maximum slope on one side of the optical resonance to obtain the maximum optical readout. We also use a proportional-integral-derivative (PID) controller to stabilize the laser wavelength. When the laser wavelength is locked, the mechanical mode spectrum of the microdisk as well as the ultrasound response spectrum can be measured with the ESA and VNA. An ultrasonic transducer with a center frequency of 1~MHz is positioned approximately 1~cm above the microdisk. The ultrasound pressure from 30~kHz to 20~MHz generated by the ultrasonic transducer is calibrated using the method introduced in our previous work \cite{microtoroid}, after considering the ultrasound attenuation in the air.

The acousto-optic interaction includes both dispersive and dissipative coupling. First, ultrasound causes a variation in the microdisk circumference, which can shift the resonance wavelengths of the optical modes. The transmitted optical power change in this case is due to the dispersive coupling. Second, owing to their large spatial overlap with ultrasound, the flapping modes of the microdisk are excited which vary the gap between the tapered fiber and the microdisk. This gap variation leads to periodic changes in the coupling strength, i.e., periodic changes in the transmitted optical power. By locking the laser wavelength on one side of the optical resonance, both dispersive and dissipative interactions can be measured by the change in the transmitted optical power. 

\section{Results}

\begin{figure}[h!]
\centering\includegraphics[width=8cm]{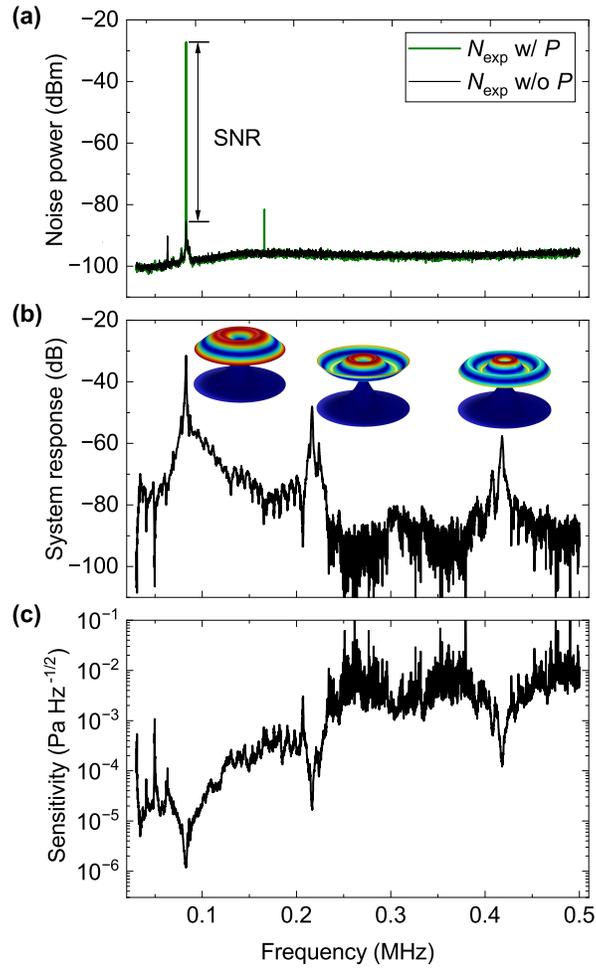}
\caption{(a) Noise power spectrum (black curve) and the response of the microdisk (green curve) driven by ultrasound at 83~kHz, with a SNR of 58.73~dB. (b) System response of the microdisk versus the ultrasound frequency. The inset shows the simulated displacement distributions of the high order flapping modes. (c) Derived ultrasound sensitivity spectrum of the microdisk.}

\end{figure}

In the experimental optimization of the ultrasound sensitivity, we obtain the best sensitivity of 1.18 ~\textmu Pa Hz$^{-1/2}$ using a microdisk with a radius of 300~\textmu m and a thickness of 2~\textmu m. The noise power spectrum obtained by the ESA without the ultrasound signal is shown in the black curve in Fig. 3(a). A mechanical resonance peak at $\Omega/2\pi$=83~kHz is observed, which corresponds to the second-order flapping mode of the microdisk. The first-order flapping mode is around 20~kHz, which is not shown here, as it is out of the frequency range in which the ultrasound pressure can be calibrated using our hydrophone. The fitted mechanical $Q$ factor is about 140. We then apply a sinusoidal voltage at this frequency to drive the ultrasonic transducer, and measure the ultrasound response of the microdisk, as shown in the green curve in Fig. 3(a). The signal-to-noise ratio (SNR) at the peak is 58.73~dB. The sensitivity at 83~kHz is given by
\begin{equation}
P_{\rm{min}}(\Omega) = P_{\rm{applied}}(\Omega)\sqrt{\frac{1}{\rm{SNR}}\frac{1}{\Delta f}}. 
\end{equation}
\emph{P}$_{\rm{applied}}$ is the ultrasound pressure reaching the microdisk at 83~kHz, which is calibrated to be 6.12~mPa. $\Delta f$ represents the resolution bandwidth of the ESA, which is 30~Hz. Therefore, the sensitivity at 83~kHz can be calculated as 1.29~\textmu Pa Hz$^{-1/2}$. In Fig. 3(a) we also observe a second-order mechanical sideband of the signal at 166~kHz, which originates from the nonlinear response of the optical readout mechanism \cite{microtoroid}. The system response of the microdisk is then measured by sweeping the frequency of the applied ultrasonic wave using the VNA. Figure 3(b) shows the response spectrum of the microdisk to ultrasound at different frequencies. In addition to the second-order flapping mode, we also observe two response peaks corresponding to the third- and fourth-order flapping modes at 216.6~kHz and 418.4~kHz respectively. Although they are not prominently shown in the noise power spectrum (i.e., they do not reach the thermal-noise-dominant regime), we can still observe them in the system response on account of the strong response of flapping modes to ultrasound. The sensitivity at different frequencies can be calculated as
\begin{equation}
P_{\rm{min}}(\omega) = P_{\rm{min}}(\Omega)\frac{P_{\rm{applied}}(\omega)}{P_{\rm{applied}}(\Omega)}\sqrt{\frac{N(\omega)}{N(\Omega)}\frac{S(\Omega)}{S(\omega)}},
\end{equation}
where \emph{P}$_{\rm{applied}}$($\omega$) is the applied ultrasound pressure to the microdisk at different frequencies. Using the sensitivity at 83~kHz obtained above, as well as the system response \emph{S}($\omega$) and the noise power spectral density \emph{N}($\omega$), the sensitivity spectrum of this microdisk can be derived for the entire frequency range. This sensitivity describes the lowest detectable ultrasound pressure of the microdisk, for a resolution bandwidth of 1~Hz when the SNR is 1. Figure 3(c) shows the sensitivity spectrum in the range of 0.03-0.5~MHz. A peak sensitivity of 1.18~\textmu Pa Hz$^{-1/2}$ is obtained at 82.6~kHz, which is quite close to the theoretical value of 751~nPa Hz$^{-1/2}$ obtained from Eq. (1), using the pressure difference of 4.7, and the spatial overlap of 0.16.

\begin{figure}[h!]
\centering\includegraphics[width=13cm]{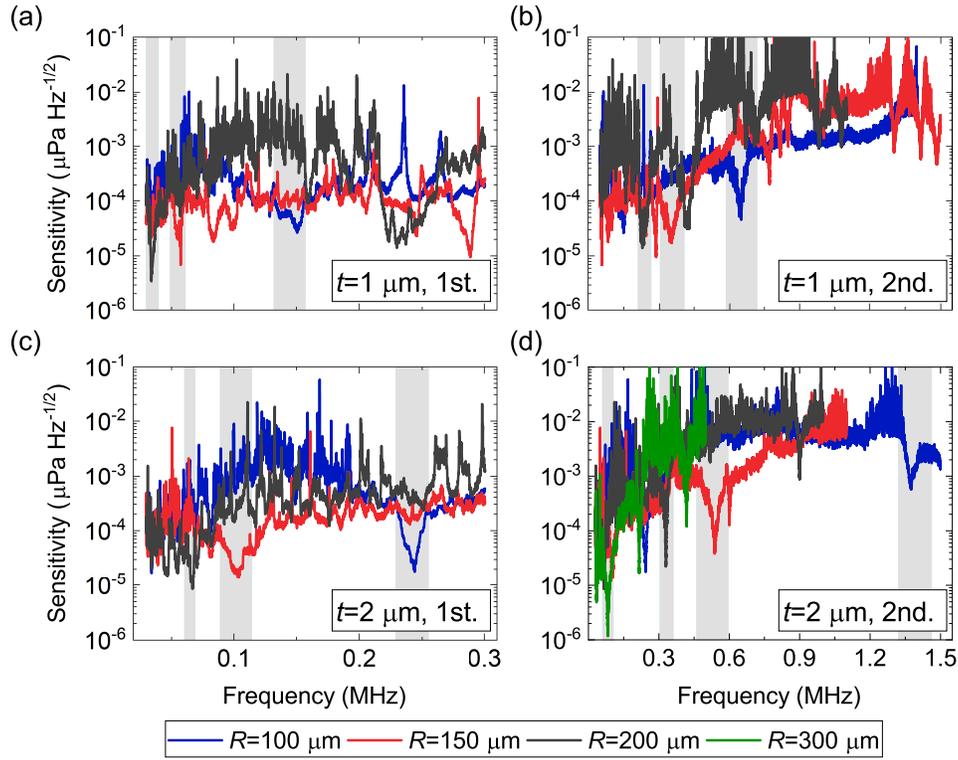}
\caption{(a) and (b) Sensitivity spectra of the 1-\textmu m-thick microdisks with different radii. (c) and (d) Sensitivity spectra of the 2-\textmu m-thick microdisks with different radii. The red, blue, black and green curves respectively represent the microdisks with radii of 100, 150, 200, and 300~\textmu m. The shaded areas highlight the first-order ((a) and (c)) and second-order ((b) and (d)) flapping modes.}
\end{figure}

\begin{figure}[h!]
\centering\includegraphics[width=8cm]{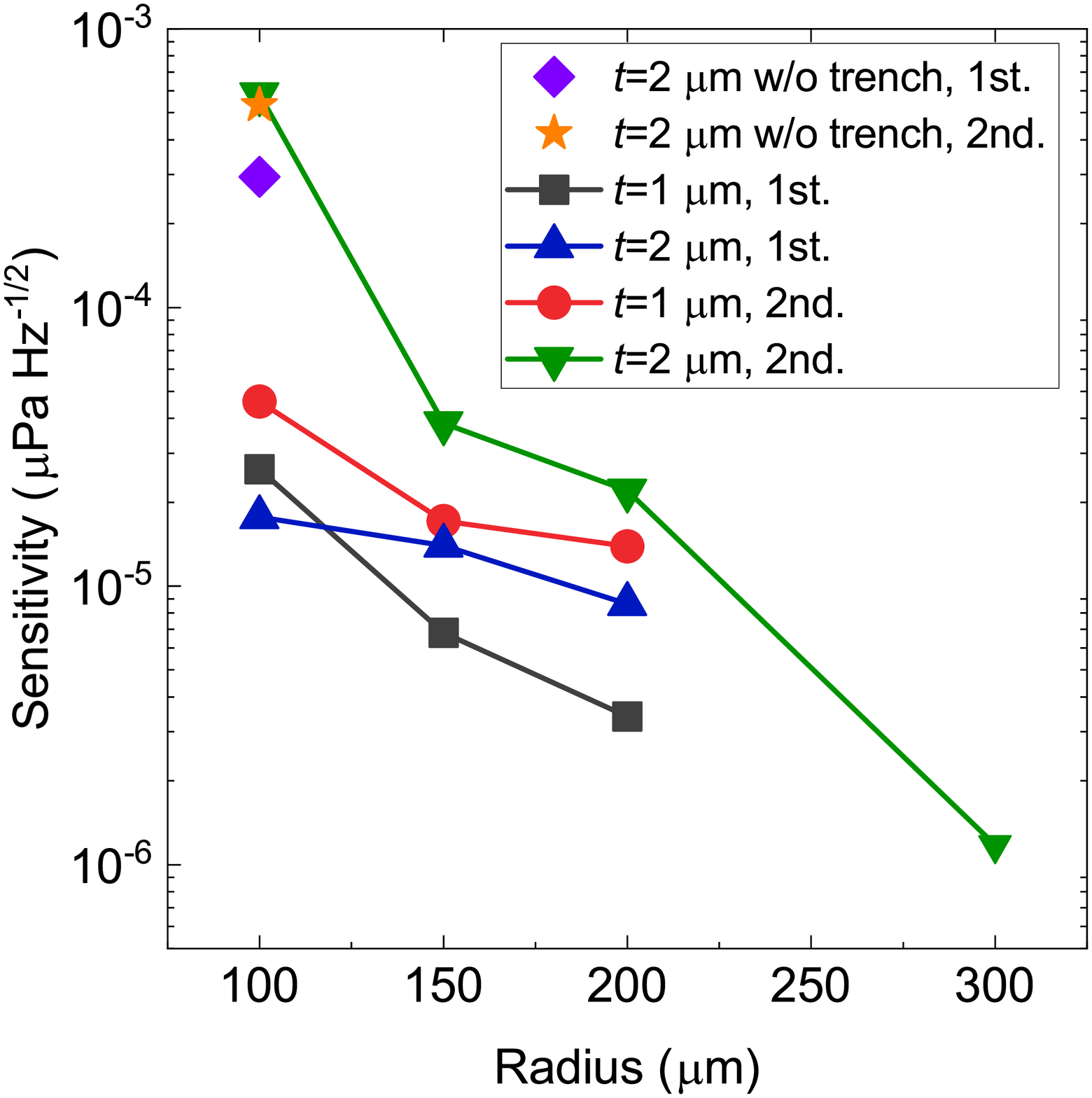}
\caption{Sensitivities at the flapping modes of microdisks with different thicknesses and radii. The black squares and blue triangles represent the sensitivities at the first-order flapping mode with thicknesses of 1~\textmu m and 2~\textmu m, respectively. The red circles and green inverted triangles represent the sensitivities at the second-order flapping mode with thicknesses of 1~\textmu m and 2~\textmu m, respectively. The purple rhombus and yellow pentagram represent the sensitivities at the first and second-order flapping modes of a microdisk with a thickness of 2~\textmu m and a radius of 100~\textmu m, but without the trench structure.}
\end{figure}

To verify the theoretical results calculated above, we measure the sensitivities of multiple microdisks of different radii and thicknesses. Figures 4(a) and 4(b) show the pressure sensitivity spectra near frequencies around the first- and second-order flapping modes for 1~\textmu m-thick microdisks with different radii. Figures 4(c) and 4(d) show the data for microdisks with 2~\textmu m thickness. The red, blue, black, and green curves respectively represent the microdisks with radii of 100, 150, 200, and 300~\textmu m, for both 1~\textmu m and 2~\textmu m thicknesses. The microdisks with a thickness of 1~\textmu m and a radius of 300~\textmu m have been cracked during the XeF$_2$ dry etching process owing to the material stress. The mechanical resonances of microdisks can significantly enhance the response to ultrasound, and therefore many dips at the mechanical modes are observed throughout the pressure sensitivity spectra. The widths of the dips which represent the detection bandwidths are determined by the thermal-noise-limited frequency range, and can be increased by using microdisks with higher optical $Q$ factors. The flapping modes are more sensitive to the ultrasound and will generally have a better sensitivity. The shaded areas highlight the sensitivities around the first-order (Figs. 4(a) and 4(c)) and second-order (Figs. 4(b) and 4(d)) flapping modes. 

From Fig. 4, it can be observed that at a certain thickness, the mechanical resonance frequency decreases and the sensitivity improves with increasing the radius, regardless of the first- or second-order flapping mode. At the same radius, the first- and second-order flapping modes have lower frequencies for thinner microdisks, which agree well with the theoretical result. For a more explicit comparison, we summarize the minimum sensitivity at the first- and second-order flapping modes in Fig. 5. It shows that the sensitivities for both the first- and second-order flapping modes improve with increasing the radius and decreasing the thickness of the microdisk. Only two points are anomalous, that is, the sensitivity of a 1~\textmu m-thick microdisk with a radius of 100~\textmu m is worse than that of a 2~\textmu m-thick microdisk. This could attribute to their different mechanical $Q$ factors and pressure differences. We also measure a microdisk with a thickness of 2~\textmu m and a radius of 100~\textmu m without a trench structure, whose sensitivities at the first- and second-order flapping modes are shown in the purple rhombus and yellow pentagram in Fig. 5. Comparing these two values with those of a same-sized microdisk with a trench structure (green inverted triangles and blue triangles) shows that the trench structure significantly improves the sensitivity of the first-order flapping mode, but does not affect the sensitivity at the second-order flapping mode. This is consistent with the theoretical results in Figs. 1(c)-(f).

\section{Conclusion and discussion}
In conclusion, we have theoretically and experimentally studied the sensitivity of the microdisk ultrasound sensor in the thermal-noise-dominant regime. Our theoretical results show that, the ultrasound sensitivity can be optimized by increasing the radius and decreasing the thickness of the microdisk, but is also affected by the pressure difference between the upper and lower surfaces of the disk. We also find that using a trench structure around the microdisk can increase the pressure difference and therefore improve the sensitivity. We have then fabricated multiple microdisks of different radii (100~\textmu m, 150~\textmu m, 200~\textmu m, and 300~\textmu m) and thicknesses (1~\textmu m and 2~\textmu m) with trench structure, and characterized their ultrasound sensitivities. Our experimental results have verified that better sensitivities can be achieved using microdisks with larger radius and thinner thickness. After a series of optimization, a peak sensitivity of 1.18~\textmu Pa Hz$^{-1/2}$ has been achieved at 82.6~kHz in the air, using a microdisk with a radius of 300~\textmu m and a thickness of 2~\textmu m. This sensitivity is to our knowledge the record sensitivity among the cavity optomechanical ultrasound sensors so far.

Such high ultrasound sensitivity can improve the detection range of ultrasound detection technology, such as increasing the detection depth of photoacoustic imaging, and is particularly helpful for the ultrasound detection in the air, which has been proved challenging due to the impedance mismatch. Furthermore, the fabrication process of microdisks is relatively mature, which allows mass production on a silicon chip. The bandwidth of the ultrasound sensor can be further improved by using microcavities with higher optical $Q$ factors and optomechanical coupling coefficients. Integrated waveguide-coupled microcavities and on-chip arrays of ultrasound sensors, such as silicon nitride and silicon microcavity sensors, can be designed in the future for practical application such as photoacoustic imaging and spectroscopy.

\begin{backmatter}
\bmsection{Funding}
National Key Research and Development Program of China (2021YFA1400700); National Natural Science Foundation of China (NSFC) (62222515, 91950118, 12174438, 11934019); the basic frontier science research program of Chinese Academy of Sciences (ZDBS-LY-JSC003).

\bmsection{Acknowledgments}
The authors thank Prof. Yun-Feng Xiao for the help with the calibration of the ultrasound pressure.

\bmsection{Disclosures}
The authors declare no conflicts of interest.

\bmsection{Data Availability}
Data underlying the results presented in this paper are not publicly available at this time but may be obtained from the authors upon reasonable request.

\end{backmatter}



\bibliography{References}
\end{document}